\documentclass[%
 reprint,
 amsmath,amssymb,
 aps,prd,
]{revtex4-2}

\usepackage{graphicx}
\usepackage{dcolumn}
\usepackage{bm}
\usepackage{hyperref}
\usepackage{lipsum}
\usepackage{amsmath}
\usepackage{amssymb}
\usepackage{amsfonts}
\usepackage{xcolor}
\usepackage{academicons}

\newcommand{\g}{\textit{g}}
\newcommand{\Lag}{\mathcal{L}}
\newcommand{\Mpl}{M_{\rm Pl}}

\def\pmin{\phi_{\rm min}}

\begin{document}

\preprint{APS/123-QED}

\title{Signatures of Solar Chameleons in the Earth's Magnetic Field}

\author{Tanmoy Kumar}
 \email{kumartanmoy1998@gmail.com}
\author{Sourov Roy}%
 \email{tpsr@iacs.res.in}
\affiliation{School of Physical Sciences, Indian Association for the Cultivation of Science,\\ 2A \& 2B Raja S.C. Mullick Road, Jadavpur, Kolkata 700032, India
}%

\date{\today}

\begin{abstract}
Chameleon dark energy models are a popular alternative to the standard cosmological constant model. These models consist of a new light degree of freedom, called chameleon, with a density dependent mass and a non-trivial coupling to both matter and photons. Owing to these couplings, chameleons can be produced inside the sun. However due to their density dependent mass, the chameleons produced in the solar core are screened and cannot escape whereas those produced outside the solar core, such as in the \textit{tachocline} region with energies of the order of few a keV, can escape from the sun and travel all the way towards Earth. Hence the Earth is expected to receive a flux of \textit{solar chameleons}. In this work we propose a \textit{light shining through wall} (LSW) type of experiment in which the Earth itself acts as a wall. Both photons and chameleons are incident on the light side of the Earth. While all the photons are stopped by the Earth, only a fraction of the chameleons are stopped by the earth due to screening. Those chameleons which are not screened by the earth pass directly through the Earth and exit the night side. Here these chameleons interact with the geomagnetic field and convert into X-ray photons. A space based X-ray telescope orbiting the Earth can detect these X-ray photons, while passing through the night side, thereby acting as a detector in this LSW type experiment. We show that such a kind of setup can be complementary to other terrestrial experiments looking for chameleons.
\end{abstract}

\maketitle


\section{Introduction}
\label{sec:intro} 
Observations suggest our universe is currently undergoing a phase of accelerated expansion~\citep{SupernovaSearchTeam:1998fmf,SupernovaCosmologyProject:1998vns}. This accelerated expansion is thought to be driven by a mysterious quantity dubbed \textit{dark energy} (DE). The recent measurements from the \textit{Planck} mission~\cite{Planck:2013pxb} suggest that almost 70\% of our universe's energy budget is made up of this mysterious \textit{dark energy} and yet we still do not know what it is~\citep{Frieman:2008sn,Joyce:2014kja,Huterer:2017buf}. The simplest DE candidate is a cosmological constant (CC) which arises due to the collective zero point energy of all the quantum fields existing in our universe. However this model of DE has a serious disadvantage in that the calculated value of this zero point energy from quantum field theory is orders of magnitude greater than the required value for explaining the current acceleration. This mismatch is referred to as the \textit{cosmological constant problem}~\citep{Sola:2013gha,Burgess:2013ara,Padilla:2015aaa}, and it is one of the most perplexing problems in modern physics. Several alternative models have been proposed to explain DE in order to avoid the cosmological constant problem. A common feature of such models is the presence of additional degrees of freedom which drive the accelerated cosmic expansion. Such models predict a wide array of novel astrophysical and cosmological phenomena which are being or will be explored by present or future cosmological surveys~\citep{SimonsObservatory:2018koc,2019BAAS...51g.147L,CMB-S4:2016ple,Abazajian:2019tiv,EUCLID:2011zbd,Amendola:2016saw,DESI:2016fyo,LSSTDarkEnergyScience:2012kar,Spergel:2013tha,Saltas:2022ybg}.

One popular class of such alternative DE models are the quintessence models. Such a kind of model was first proposed in~\citep{Ratra:1987rm}. These models propose that the accelerated cosmic expansion is due to a scalar field rolling down a flat potential. In order for the scalar field to evolve  cosmologically, its mass must be of the order of the present Hubble scale $H_0$, implying that it must be an ultralight field. Now generically such a scalar field can also have a small coupling with matter. However such a matter coupling would lead to a violation of the equivalence principle (EP) and result in the existence of a \textit{fifth force} and the existence of such a fifth force is strongly constrained by solar system tests of GR~\citep{Will:2005va,Sakstein:2017pqi}. This problem is avoided if there exists some form of a \textit{screening mechanism} in the theory which dynamically suppresses fifth-force in the solar system~\cite{Khoury:2010xi,Brax:2013ida,Sakstein:2018fwz,Baker:2019gxo}. 
The \textit{chameleon effect}, first proposed in~\cite{Khoury:2003aq,Khoury:2003rn}, is one such screening mechanism. A scalar field exhibiting the chameleon effect is called a chameleon field or simply chameleon. The primary feature of a theory exhibiting the chameleon effect is the existence of a coupling of the chameleon field with matter that leads to a unique density dependent mass of the scalar field~\cite{Burrage:2016bwy,Burrage:2017qrf}. Typically in these models the mass of the chameleon field increases with increasing matter density and vice-versa. Thus, in regions of high matter density, such as the Earth, the mass of the chameleon field can be sufficiently large for it to evade fifth-force search experiments whereas in the cosmic vacuum, where the density is almost zero, it will have a very small mass as required for explaining the cosmic expansion.

The coupling of the chameleon field with matter almost automatically implies the existence of a coupling with photons~\cite{Brax:2009ey}. This chameleon-photon coupling has many interesting phenomenological consequences and it has been used to search for DE chameleons both directly and indirectly~\cite{Burrage:2007ew,Burrage:2008ii,GammeV:2008cqp,Steffen:2010ze,Schelpe:2010he,CAST:2015npk}. In this work we focus on one such phenomenon due to the chameleon-photon coupling. Chameleon theories, being screened theories, imply that chameleon particle production inside the core of the sun is highly suppressed due to the enormously high density of the environment. Nevertheless due to the presence of the chameleon-photon coupling, chameleon particles can still be produced inside the Sun in regions outside the core where there exists strongly magnetized plasma. In particular, the solar tachocline, which is the region of transition from the radiative interior to the convective outer zone, is expected to host extremely strong magnetic fields. In this region chameleon particles can be produced from thermal photons via a Primakoff-like process. The produced chameleons can be relativistic and, owing to their weak interaction with matter, can have a large mean free path. This in turn would enable the chameleon particles to escape from the sun resulting in a solar chameleon flux. This possibility in turn opens up a new avenue in the detection of DE and it has been explored in several works~\cite{Brax:2010xq,Brax:2011wp,Brax:2015fya,CAST:2015npk,Vagnozzi:2021quy}.

In this work we too explore the above avenue but in a different setup. The solar chameleons escaping the Sun spread out in all directions. Hence all the planets, including the Earth, is expected to receive a flux of solar chameleons alongside photons. Whereas all the photons are blocked by the Earth, a fraction of the chameleon flux can travel right through the Earth and exit the night side of the Earth where the solar photon background would essentially be zero. We show that these exiting chameleons can convert into photons, via the same Primakoff-like process, while travelling through the Earth's magnetic field. We further show that for certain chameleon models these photons, which would have energies of the order of a few keV, would give rise to X-ray signals which can be detected by future X-ray telescopes orbiting the Earth. In such cases we find that non-observation of an X-ray signal will set constraints on the chameleon-photon coupling which are complementary to the existing constraints.

The rest of this paper is organized as follows. In Sec.~\ref{sec:chameleon} we give a brief outline of chameleon scalar field models. In Sec.~\ref{sec:chameleon_photon} we elaborate on the chameleon-photon coupling, discussing its phenomenological consequences, most importantly the phenomenon of chameleon-photon oscillation in an external magnetic field. The production of chameleons inside the Sun via the \textit{chameleonic Primakoff-like effect} is discussed in Sec.~\ref{sec:solar_chameleon} using which we calculate the solar chameleon flux. In Sec.~\ref{sec:xray} we outline how chameleon photon oscillation results in the production of a flux of X-ray photons in the Earth's atmosphere. We also explore the parameter space of different chameleon models and find the ones that can result in a detectable X-ray signal in a future X-ray observatory. We summarize and conclude in Sec.~\ref{sec:conclu}.

\section{Chameleon Model}
\label{sec:chameleon} 
The chameleon model~\cite{Khoury:2003aq,Khoury:2003rn} is a scalar field model of DE. It consists of a scalar field, which we call the chameleon field or simply chameleon, $\phi$, whose main features are - (a) a runaway potential (typically taken to be a power law type potential) and (b) a non-trivial coupling to matter. The net result of these two features is an effective potential for the chameleon which depends on the matter density of the surrounding environment leading to an environment dependent mass of the chameleon. This environment dependent mass is the primary feature of this model.

In particular, the chameleon model introduces an additional scalar field, $\phi$, whose behavior is governed by the Lagrangian~\cite{Khoury:2003aq}, 
\begin{multline}
    \Lag \supset \sqrt{-\g} \left[ \frac{M_{\rm pl}^2}{2}R(\g) - \frac{(\partial_\mu \phi)^2}{2} - V(\phi) \right]\\ + \Lag_{m}(\psi^{(i)}, \g_{\mu \nu}^{(i)})
    \label{eqn:lagrangian}
\end{multline}
where $M_{\rm Pl}$ is the reduced Planck mass, $R(\g)$ is the Ricci scalar and $\psi^{(i)}$ are the matter fields. Each matter field, $\psi^{(i)}$, couples conformally to $\phi$ through the metric
\begin{equation}
    g_{\mu \nu}^{(i)} = A^2(\phi) g_{\mu \nu}
    \label{eqn:conformal_metric}
\end{equation}
where $g_{\mu \nu}$ is the Einstein frame metric and $A(\phi)$ is the conformal coupling function. Here $V(\phi)$ is the potential of the chameleon field. 

The form of the potential $V(\phi)$ is so chosen that it is monotonically decreasing and satisfies the conditions $V,\; dV/d\phi,\; d^2 V/d \phi^2, ...\rightarrow 0$ as $\phi \rightarrow \infty$ and $V,\; dV/d\phi,\; d^2 V/d \phi^2, ...\rightarrow \infty $ as $\phi \rightarrow 0$. Out of many possible potentials that satisfy the previous criteria, in this work we shall focus on inverse power law models~\citep{Ratra:1987rm, Burrage:2016bwy} with a potential of the form
\begin{equation}
V(\phi) = {\Lambda^\prime}^4 + \frac{\Lambda^{4+n}}{\phi^n}
\label{eqn:potential}
\end{equation}
where $\Lambda$ and $\Lambda^\prime$ are some energy scales of the model. For the chameleon model to account for the observed accelerated expansion of the Universe at large scales, we require that $\Lambda^\prime = \Lambda_{\rm DE} = 2.4$ meV~\citep{Burrage:2016bwy, Burrage:2017qrf}. The parameter $\Lambda$ on the other hand is not bound by any such restrictions. Two most commonly used choices for this parameter are~\citep{elder2023constraining} are $\Lambda = \Lambda_{\rm DE}$ and $\Lambda = 1\;\mu$eV. The second choice corresponds to the region of the parameter space where solar chameleons could show up in dark matter direct detection experiments~\citep{Vagnozzi:2021quy}. In this work we shall consider the second choice of $\Lambda$. As for the parameter $n$, it can only take certain integer values as discussed later.

Coming to the conformal coupling function, again there are a wide variety of possible choices. In this work we shall assume an exponential coupling of the form
\begin{equation}
A(\phi) = e^{\beta_m \phi / \Mpl}
\label{eqn:conformal_coupling}
\end{equation}
where $\beta_m$ is the dimensionless coupling constant of chameleon with matter.

Given the potential and the coupling function, starting from the Lagrangian in Eqn.~\eqref{eqn:lagrangian} and working in the Einstein frame,we can now obtain the equation of motion of the chameleon field~\citep{Khoury:2003aq, Burrage:2017qrf}
\begin{equation}
\square \phi = \frac{d V_{\rm eff}}{d \phi}
\label{eqn:chameleon_eom}
\end{equation}
where $V_{\rm eff}$ is given by
\begin{equation}
V_{\rm eff} = V(\phi) + \rho e^{\beta_m \phi / \Mpl}
\label{eqn:V_eff}
\end{equation}
and $\rho$ is the total non-relativistic matter density in the surrounding of the chameleon field. Thus we see from Eqn.~\eqref{eqn:chameleon_eom} that the dynamics of the chameleon field is governed by an \textit{effective potential} $V_{\rm eff}$ which is explicitly dependent on the surrounding non-relativistic matter density $\rho$. 

The above effective potential has a density dependent minimum located at
\begin{equation}
\phi_{\rm min} = \left( \frac{n \Mpl \Lambda^{4+n}}{\beta_m \rho} \right)^{1/(n+1)}.
\label{eqn:phi_min}
\end{equation}
The chameleon rest mass squared, which is simply $d^2 V_{\rm eff}/d \phi^2$ evaluated at $\phi = \phi_{\rm min}$ is

\begin{equation}
 m_c^2(\rho)= e^{\frac{\beta_m \phi_{\rm min}}{\Mpl}} \frac{\beta_m \rho}{\Mpl} \left( \frac{n+1}{\phi_{\rm min}} + \frac{\beta_m}{\Mpl} \right),
\label{eqn:m_squared}
\end{equation}
which is, as mentioned previously, dependent on the non-relativistic matter density $\rho$.
An important point to note from Eqns.~\eqref{eqn:V_eff}, \eqref{eqn:phi_min} and \eqref{eqn:m_squared} is that the nature of the dependence of $\pmin$ and $m_c$ on $\rho$ crucially depends on the parameter $n$. 

Now as we mentioned earlier, models with extra light scalar fields are severely constrained from solar system tests for the existence of a fifth force. On the other hand for a scalar field driven accelerated expansion of the universe, the scalar field must be light. Hence for the chameleon model to successfully explain DE, the chameleon mass must depend on $\rho$ in such a way that in a region of low matter density (e.g., the intergalactic space) it has an extremely low mass whereas in a region of very high matter density (e.g., inside Sun or other solar system objects) it has a very high mass so that any fifth-force mediated by this scalar particle is effectively screened~\citep{Burrage:2016bwy,Burrage:2017qrf}. These conditions are satisfied when either $n$ is a positive integer or it is an even negative integer. 

In our work we shall consider $n$ to be a free parameter and show our results for different values of $n$.

\section{Chameleon-Photon Mixing}
\label{sec:chameleon_photon} 
Apart from its coupling to matter, the chameleon field, in general, can also couple to photons. Following~\cite{Brax:2010xq}, we assume the chameleon-photon interaction term to be of the form
\begin{equation}
    \Lag_{\rm int} = \sqrt{-\g} \frac{e^{\beta_\gamma \phi / \Mpl}}{4} F_{\mu \nu} F^{\mu \nu}
    \label{eqn:chameleon_photon_coupling}
\end{equation}
where $\beta_\gamma$ is a dimensionless coupling constant which we consider to be a parameter of the model. Here $F_{\mu \nu}$ is the electromagnetic field strength tensor. The dimensionless chameleon-photon coupling has already been constrained from different astrophysical observations and ground based experiments (see~\cite{elder2023constraining} for a summary of existing constraints). 

An immediate consequence of this coupling is that the chameleon mass is also affected by the energy density of the electromagnetic field in the environment. Thus the effective chameleon mass is a function of both the matter density, $\rho$ and the electromagnetic energy density $B^2/2$, where $B$ is the magnetic field in the environment. Typically the contribution of the electromagnetic field to the total energy density is orders of magnitude smaller than the contribution from ordinary matter density and hence throughout our work we ignore the electromagnetic contribution. A second consequence of this chameleon-photon coupling is that the chameleon mixes with the photon. This mixing arises due to the term 
\begin{equation}
    \Lag_{1} = \frac{\beta_\gamma}{\Mpl} \phi B^2
    \label{eqn:L_gamma}
\end{equation}
in the interaction Lagrangian~\eqref{eqn:chameleon_photon_coupling}. In presence of an external magnetic field, the mixing term results in a beam of chameleons oscillating into photons and vice-versa. This chameleon-photon oscillation in external magnetic field is not only responsible for the production of chameleons inside the Sun but we shall also employ this same mechanism to detect these chameleons. Hence we shall first outline the dynamics of a chameleon oscillating into a photon in an external magnetic field.

Assuming a beam of chameleons propagating along the $z$ axis in a homogeneous and transverse external magnetic field of strength $B$ which is aligned in a fixed direction, the equation of motion (EoM) of the chameleon and the electromagnetic fields are~\citep{Burrage:2007ew}
\begin{equation}
\left[ \omega^2 + \partial^2_z + 
\begin{bmatrix}
-\omega^2_p & 0 & 0 \\
0 & -\omega^2_p & \frac{\beta_\gamma B \omega}{\Mpl} \\
0 & \frac{\beta_\gamma B \omega}{\Mpl} & -m_c^2
\end{bmatrix}
\right]
\begin{bmatrix}
A_{\parallel} \\
A_{\bot} \\
\phi
\end{bmatrix} = 0
\label{eqn:osc_eom}
\end{equation}
where $\omega$ is the energy of the chameleons/photons and $\omega_p^2 = 4 \pi \alpha n_e / m_e$ is the photon plasma mass in the medium and $n_e$ is the electron density in the medium through which the chameleons are propagating. Here $A_{\parallel}$ and $A_{\bot}$ are the parallel and perpendicular components of the photon field with respect to the external magnetic field. The above EoM is similar to the photon-axion EoM in external magnetic field excepting a crucial difference - the axion couples to the polarization parallel to the external magnetic field ($A_\parallel$)~\citep{Raffelt:1987im} whereas the chameleon couples to the polarization perpendicular to the external magnetic field ($A_\bot$). This is a crucial difference when it comes to differentiating whether a potential signal originated due to axion conversion into photon or a chameleon conversion into photon.

Similar to the case of axion as shown in~\citep{Raffelt:1987im}, we first linearize Eqn.~\eqref{eqn:osc_eom} to get the linearized EoM. Since $A_\parallel$ evolves independently we can ignore it. We then solve the linearized EoM with the initial conditions - $A_\bot = 0$, $\phi = 1$ and obtain the chameleon to photon oscillation probability after travelling a distance $z$ as
\begin{equation}
P_{c \leftrightarrow \gamma} (z) = {\rm sin}^2 \theta {\rm sin}^2 \left( \frac{\Delta (z)}{2{\rm cos} 2\theta} \right)
\label{eqn:osc_prob}
\end{equation}
where
\begin{equation}
{\rm tan} 2\theta = \frac{2 \omega \beta_\gamma B}{\Mpl (\omega_p^2 - m_c^2)},
\end{equation}
and $\Delta(z) = (\omega_p^2 - m_c^2) z / 2\omega$.

In the calculation above we have neglected two effects. Firstly, in presence of an external magnetic field, the vacuum becomes birefringent. This effect is manifested in the above EoM as an additional mass term $\Delta_{\rm QED}$ of both the photon polarization states. Additionally in presence of an external magnetic field, the medium through which the photons propagate also become birefringent (Cotton-Mouton effect) which also results in a mass term for both of the photon polarization states. However, for the magnetic field strength values considered here, the contribution of both of these effects to the photon's effective mass is negligible as compared to the plasma mass in the medium. Hence these effects can be safely neglected.

With the above chameleon to photon oscillation probability we now show how this mechanism leads to the production of chameleons inside the sun.

\section{Solar Chameleons: Chameleonic Primakoff-like Effect}
\label{sec:solar_chameleon} 

\begin{figure*}[ht!]
\centering
\includegraphics[width=10cm]{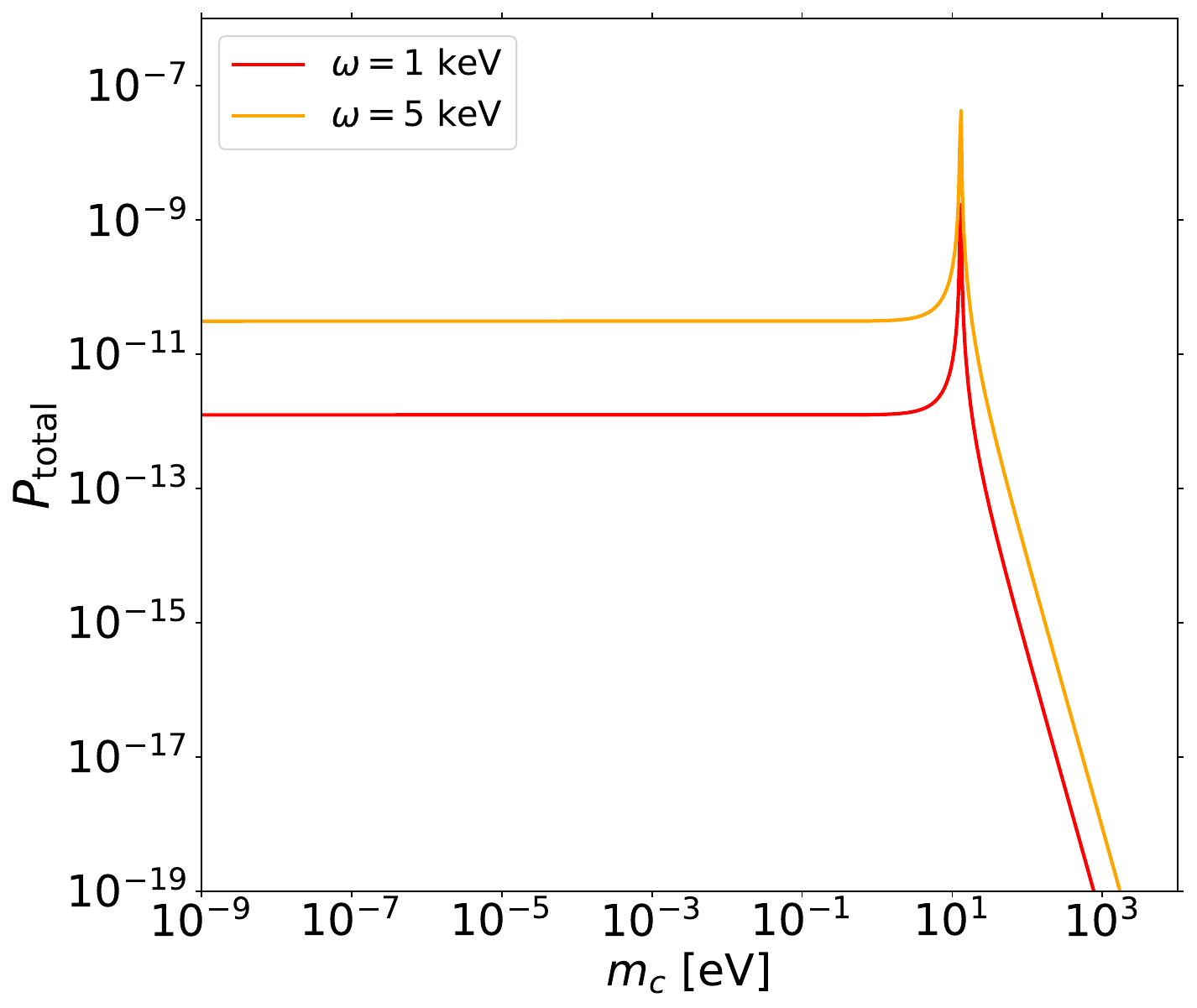}
\caption{Total oscillation probability $P_{\rm total}$ of a thermal photon into chameleon in the solar tachocline as a function of the effective chameleon mass for two different values of thermal photon energy $\omega = 1$ keV and $\omega = 5$ keV. The values of the chameleon couplings are taken to be - $\beta_\gamma = 10^{13}$ and $\beta_m = 10^{2}$. The strength of the magnetic field in the tachocline is chosen to be $B = 30$ T. The sharp peak in the oscillation probability is due to resonance where the chameleon mass in the tachocline becomes equal to the photon plasma mass in the tachocline.
}
\label{fig:P_total}
\end{figure*}

Chameleons can be produced inside the Sun thanks to their coupling with the photon. From the form of the chameleon-photon coupling, one can naively expect that just like axions, chameleons can be copiously produced via Primakoff effect, Compton scattering and bremsstrahlung emission. However, thanks to the density dependent mass of the chameleon, inside the Sun where densities are extremely high, the high mass of the chameleons largely suppress the above three processes in the solar interiors thereby preventing efficient production of chameleons. As a result the Sun is effectively screened~\citep{Burrage:2016bwy,Burrage:2017qrf}, which is in fact an essential requirement of the chameleon model in order to escape the bounds from fifth-force searches. Nevertheless chameleons can still be produced inside the Sun via the chameleonic Primakoff-like effect~\citep{Brax:2010xq,Vagnozzi:2021quy}.

Inside the Sun, there exists strong magnetic fields. The thermal photons present inside the Sun, while traversing through these strong magnetic fields can oscillate into chameleons. Unlike the photons which have a small mean free path in the fully ionized interior of the Sun, the chameleons, by virtue of their weak interaction with matter, have much longer mean free paths. 
Hence chameleons produced close to the surface can escape the Sun thereby contributing to the solar chameleon flux.

In this work following~\citep{Brax:2010xq,Brax:2011wp,Brax:2015fya}, we consider chameleon production from photons in 
the solar \textit{tachocline} region. The tachocline region is a thin shell of thickness 
$0.05 R_{\odot}$ located at a radius of $R_{\rm tacho} = 0.7 R_{\odot}$, where $R_{\odot}$ is the radius of the Sun. It is generally believed that this tachocline is the source of the solar magnetic field and it is accepted that this region has a magnetic field of strength $B = 20-50$ T~\citep{Couvidat:2002gvk} and a density of $\rho = 0.2$ g$/$cm$^3$. For this work, we shall use the rather conservative value of $B = 30$ T. 

The thermal photon flux in the tachocline region is around $n_\gamma = 10^{21}$ cm$^{-2}$ s$^{-1}$. These photons have a plasma frequency given by $\omega_p^2 = 4 \pi \alpha n_e / m_e$ where $n_e = \rho / m_p$ assuming electrical neutrality of the solar plasma with $m_p$ being the proton mass. These photons, while passing through the tachocline, oscillate and produce chameleons. The probability of a single photon converting into chameleon over the length of one mean free path while travelling through the tachocline is
\begin{equation}
P_{\rm chameleon}(\omega) = {\rm sin}^2 \theta \left\langle {\rm sin}^2 \left( \frac{\Delta}{{\rm cos} 2\theta} \right) \right\rangle,
\label{eqn:avg_osc_probability}
\end{equation}
where we average out the chameleon production over the photon mean free path $\lambda$ in the tachocline
\begin{equation}
\left\langle {\rm sin}^2 \left( \frac{\Delta}{{\rm cos} 2\theta} \right) \right\rangle = \frac{1}{\lambda} \int_0^\lambda dz \; {\rm sin}^2 \left( \frac{\Delta(z)}{{\rm cos} 2\theta} \right).
\label{eqn:average_formula}
\end{equation}
Typically in the tachocline $\lambda = 10$ cm. For the standard values of the parameters involved, $\Delta(\lambda) \gg 1$ and $\theta \ll 1$ implying
\begin{equation}
P_{\rm chameleon}(\omega) = \frac{1}{2} \theta^2.
\label{eqn:P_chameleon}
\end{equation}
Hence the probability of producing one chameleon in one second out of one thermal photon is
\begin{equation}
P_{\rm total}(\omega) = N P_{\rm chameleon}(\omega),
\label{eqn:P_total}
\end{equation}
as $P_{\rm chameleon} \ll 1$. Here $N = \frac{1\text{s}}{\lambda}$ is the number of interactions the photon experiences inside the solar magnetic field in 1 second. 

Shown in Fig.~\ref{fig:P_total} is the probability of producing one chameleon in one second out of one thermal photon Eqn.~\eqref{eqn:P_total} in the solar tachocline as a function of chameleon mass for two different choices of thermal photon energy $\omega$. From this figure we see that for low chameleon masses $P_{\rm total} $ is constant. It sharply rises due to resonance when the chameleon mass becomes equal to the photon plasma mass in the tachocline  and then it rapidly drops for even higher chameleon masses.

Now assuming a blackbody distribution for the thermal photons 
\begin{equation}
p_\gamma (\omega) = \frac{\omega^2}{\pi^2 \bar{n}} \frac{1}{e^{\omega/T}-1}
\label{eqn:p_gamma}
\end{equation}
with temperature $T = 200$ eV and $\bar{n} = \frac{2 \zeta(3)}{\pi^2} T^3$ being the average number of photons at temperature $T$, we can obtain the chameleon spectrum produced from the solar tachocline due to the conversion of thermal photons as
\begin{equation}
\Phi_{\rm cham} (\omega) = p_\gamma (\omega) P_{\rm total}(\omega) n_\gamma.
\label{eqn:phi_chameleon}
\end{equation}
Multiplying Eqn.~\eqref{eqn:phi_chameleon} with $\omega$ we can get the energy spectrum of the solar chameleons as
\begin{equation}
l(\omega) = \omega \Phi_{\rm cham} (\omega).
\label{eqn:l_chameleon}
\end{equation}

\begin{figure*}[ht!]
\centering
\includegraphics[width=8cm]{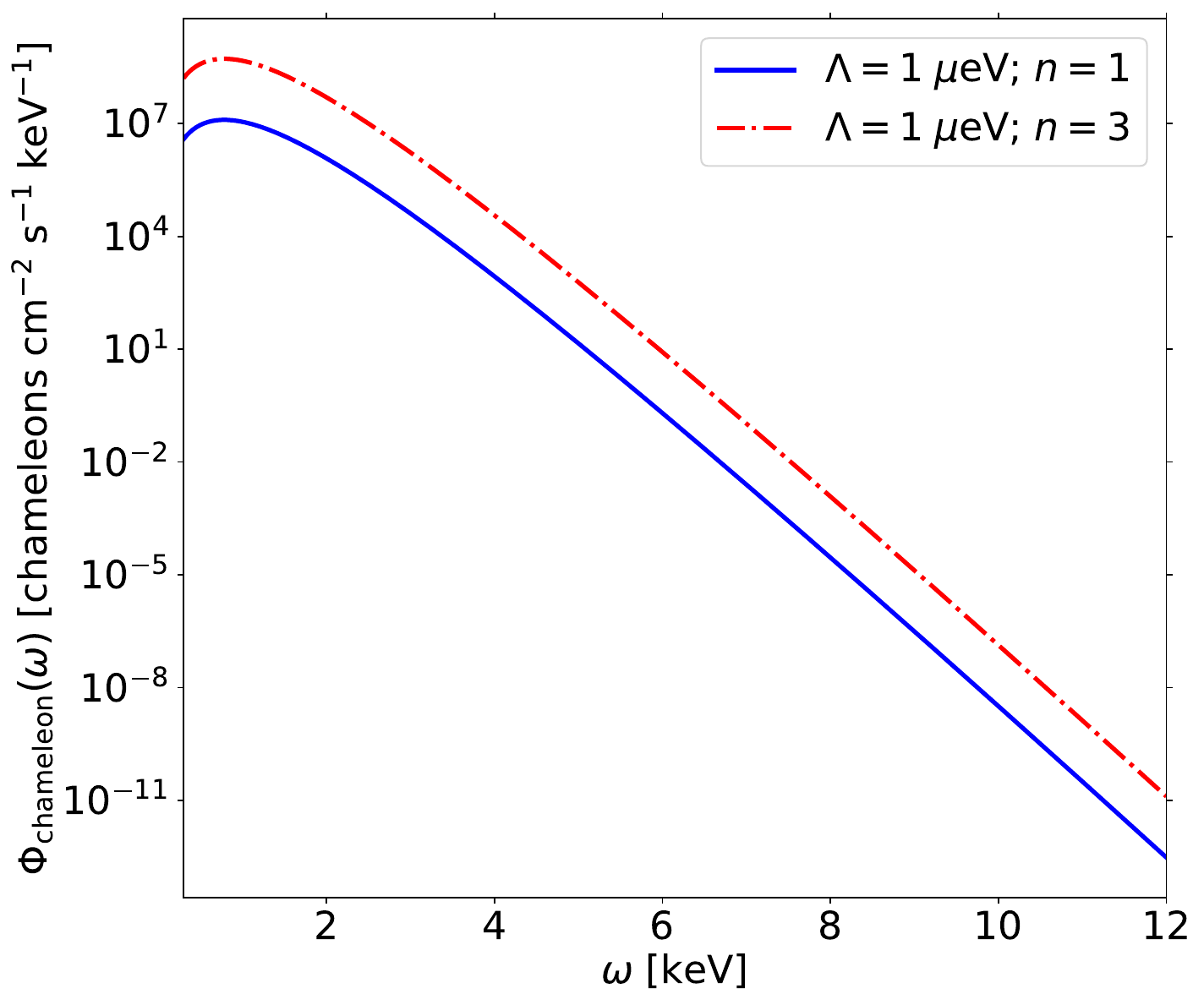}
\includegraphics[width=8cm]{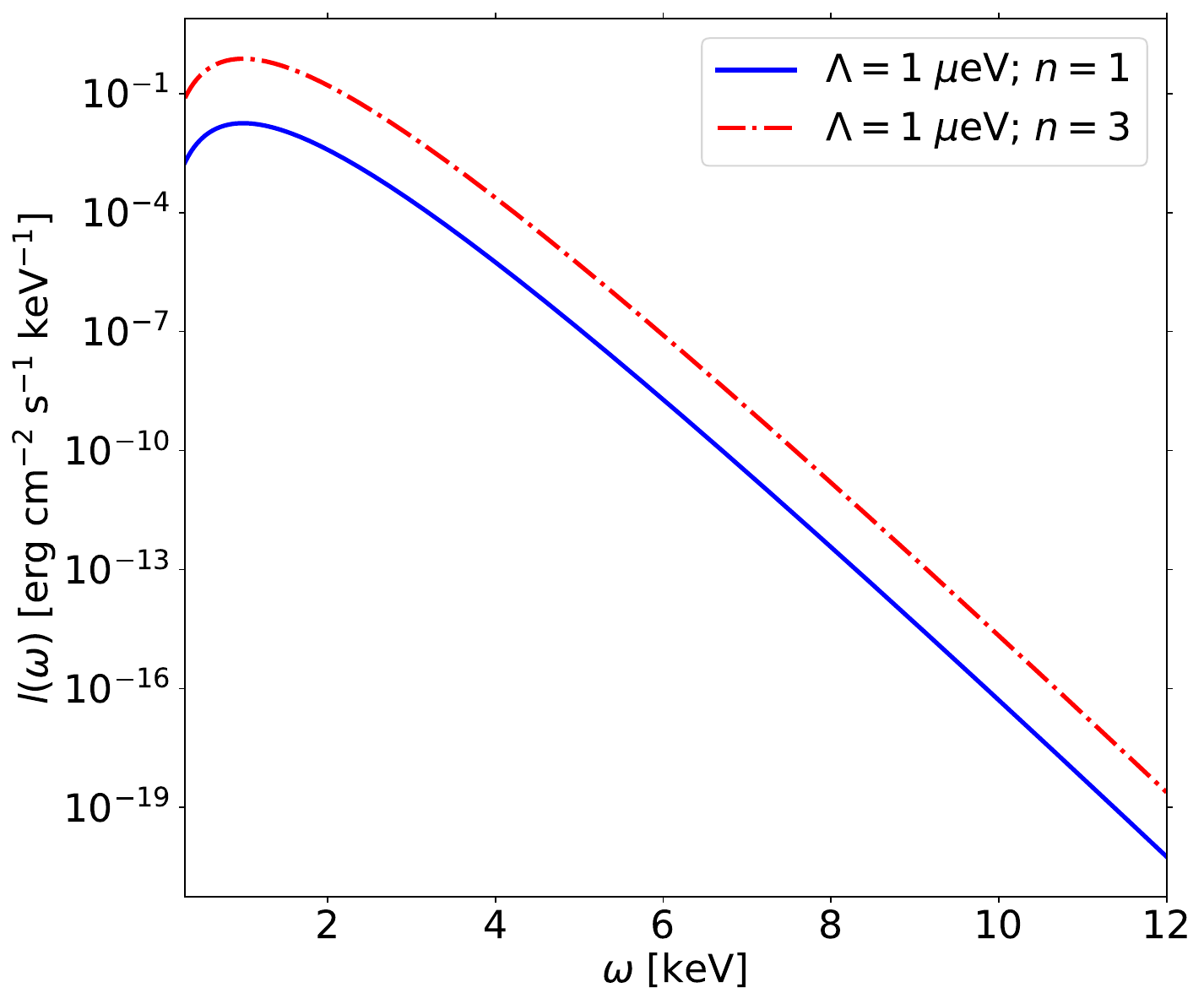}
\caption{{\it Left panel:} The differential chameleon flux produced inside the Sun due to the conversion of thermal photons in the solar tachocline for $\Lambda=1\;\mu$eV and $n=1$ (blue solid line) and $\Lambda=1\;\mu$eV and $n=3$ (red dotted-dashed line).
{\it Right panel:} The differential energy spectrum of the solar chameleons produced in the solar tachocline for $\Lambda=1\;\mu$eV and $n=1$ (blue solid line) and $\Lambda=1\;\mu$eV and $n=3$ (red dotted-dashed line). 
The values of the chameleon couplings are taken to be - $\beta_\gamma = 10^{13}$ and $\beta_m = 10^{2}$. The strength of the magnetic field in the tachocline is chosen to be $B = 30$ T.
}
\label{fig:solar_chameleon spectrum}
\end{figure*}

Shown in Fig.~\ref{fig:solar_chameleon spectrum} are the differential flux of the chameleons produced from the solar tachocline (left panel) and the energy spectrum of these produced chameleons (right panel) for two different values of $n$ namely, $n=1$ (blue solid line) and $n=3$ (red dotted-dashed line) with $\Lambda = 1\;\mu$eV in both the cases. The values of the chameleon couplings are taken as $\beta_\gamma = 10^{13}$ and $\beta_m = 10^{2}$. 
We can see from Fig.~\ref{fig:solar_chameleon spectrum} that the solar chameleon flux for the $n=1$ model (blue solid line) is lower than that for the $n=3$ model (red dotted-dashed line).
This happens because for the $n=1$ model and for the chosen value of $\beta_m$, chameleon mass in the tachocline is greater than $10$ eV and hence it is greater than the photon plasma mass in the tachocline. On the other hand for the $n=3$ model, the chameleon mass is around $1$ eV and hence it is lower than the photon plasma mass. Now from Fig.~\ref{fig:P_total} we see that for $m_c < \omega_p$ the total oscillation probability in Eqn.~\eqref{eqn:P_total} is greater than that for $m_c > \omega_p$. As a result the flux for the $n=1$ model is less than that for the $n=3$ model. Furthermore with increasing $n$, $m_c$ decreases further and goes below $1$ eV and hence for models with $n>3$, the solar chameleon flux is identical to that for the $n=3$ model.

Now these chameleons are produced over a wide range of energies. On the other hand the mass of these produced chameleons in the tachocline is $\lesssim$ keV. Using $|\vec{k}|^2 = \omega^2 - m_c^2$ where $\vec{k}$, $\omega$ and $m_c$ are the chameleon 3-momentum, energy and mass, we see that majority of the produced chameleons have $|\vec{k}|^2 \gg 0$, implying they are highly relativistic. These relativistic chameleons can escape from the solar interior and come out of the Sun resulting in a solar chameleon flux.

In the above discussion we have left out an important issue. The chameleons produced in the tachocline, while propagating outside, can convert back into photons due to inverse Primakoff effect, which in turn will get trapped. Thus a portion of the chameleon flux can be lost inside the Sun. However similar to the case of axions, we expect this phenomenon to not have much effect in the outgoing chameleon flux. 

\section{Chameleon Signatures in X-ray Observations}
\label{sec:xray}
The chameleons produced in the Sun along with the photons travel towards the Earth and are incident on the light side the Earth. Once they reach Earth, whereas all the photons are absorbed by the surface of the Earth, some of the chameleons can pass directly through the Earth while a small portion cannot. This is again due to the density dependent mass of the chameleon.

The density of the Earth increases as we move from the surface towards the core. Thus the mass of the chameleons is different at different depths. Using the preliminary reference Earth model~\citep{Dziewonski:1981xy}, we find that the mass of the chameleons at the core, where the density is highest, can vary from few $\mu$eV to tens of keV depending on the value of $\beta_m$. Hence for a given $\beta_m$ only chameleons with an energy $\omega \gg m_c (\rho_{\rm core})$ can pass through the Earth and escape into the night side of the Earth. Once outside the Earth, these chameleons travel through the geomagnetic field. This geomagnetic field will induce chameleon-photon oscillation which can result in observable X-ray signals. Before calculating the observable X-ray flux, we first calculate the chameleon-photon oscillation probability in the geomagnetic field.



The magnetic field of the Earth, referred to as the geomagnetic field, can be well approximated as a dipole field for distances less than $1000$ km above the surface. The field strength at the equator is $B_{\oplus} \simeq 3 \times 10^{-5}$ T and it falls off as $1/r^3$. However over distances $L \ll R_{\oplus}$, where $R_{\oplus}$ is the radius of the Earth, the geomagnetic field can be assumed to be fairly constant. Therefore the chameleon-photon oscillation probability in the geomagnetic field is simply the oscillation probability in a homogeneous field.
Hence the probability of the escaping chameleons to oscillate into photons after travelling a distance $L$ in the geomagnetic field is simply given by Eqn.~\eqref{eqn:osc_prob} which we rewrite as
\begin{equation}
P_{\gamma} (L) = 2 \left( \frac{\beta_\gamma B_{\oplus}}{2 \Mpl} \right)^2 \left( \frac{1 - {\rm cos}(qL)}{q^2} \right),
\label{eqn:osc_prob_Earth}
\end{equation}
where $q = | \omega_p^2 - m_c^2 | / 2\omega $ with $\omega_p$ being the photon plasma mass and $m_c$ being the chameleon mass in the Earth's atmosphere.

Hence at a fixed distance $L$ from the surface of the Earth in the night side, the expected differential photon flux due to the conversion of chameleons is
\begin{equation}
\Phi_{\gamma} (\omega) = P_\gamma (L) \frac{R^2_{\rm tacho}}{D^2_{\odot}} \Phi_{\rm cham} (\omega)
\label{eqn:xray_photon_flux}
\end{equation}
where $D_{\odot}$ is the distance from the Earth to the Sun and is equal to 1 A.U.

\begin{figure*}[ht!]
\centering
\includegraphics[width=10cm]{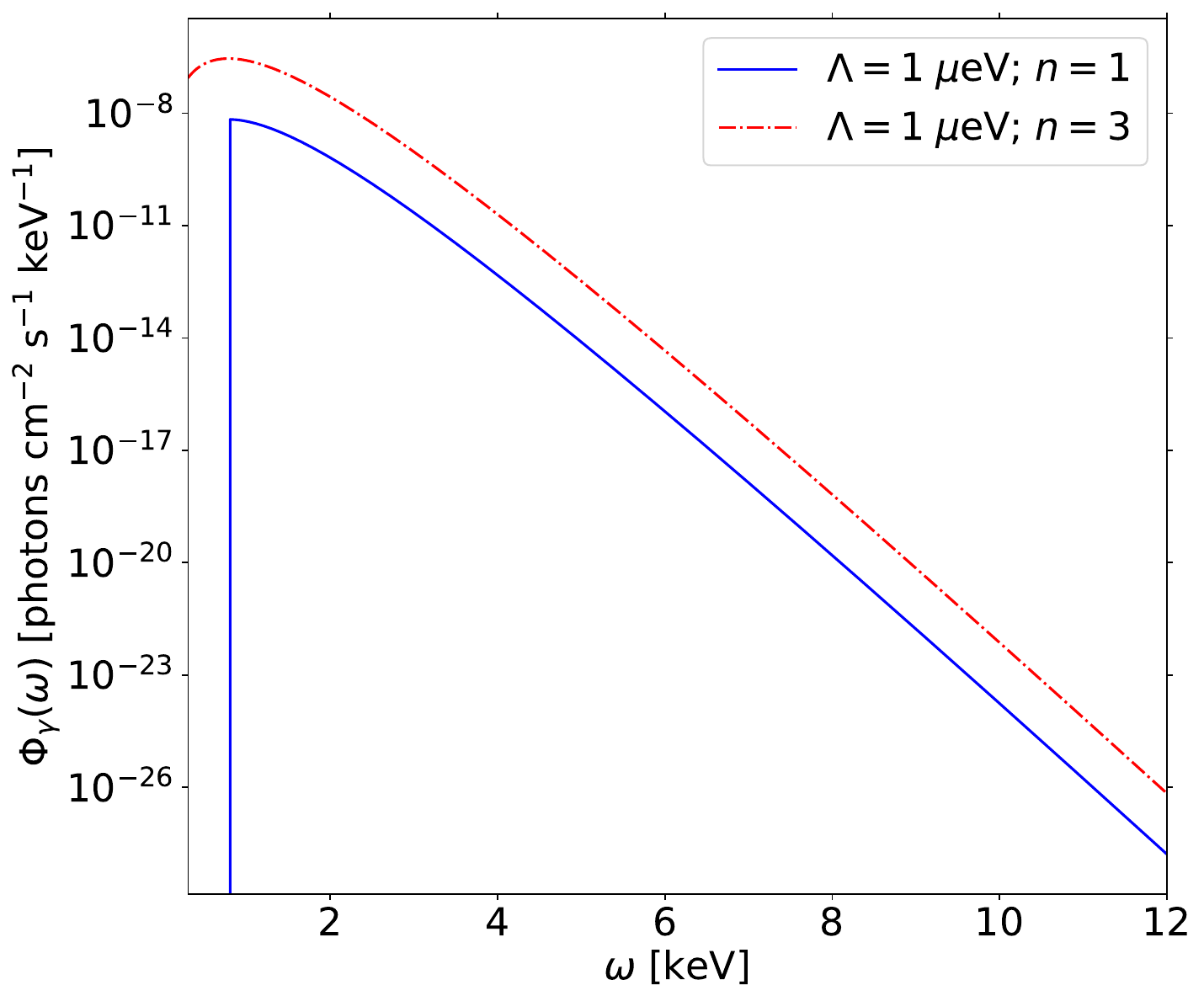}
\caption{Spectrum of the regenerated X-ray photons in the night side of the Earth at a distance of 600 km from the surface, produced due to the conversion of solar chameleons into photons in the geomagnetic field. The blue solid line and red dotted-dashed line correspond to chameleon models with $\Lambda = 1\;\mu$eV and $n=1$ and $\Lambda = 1\;\mu$eV and $n=3$. The values of chameleon-matter and the chameleon-photon coupling are chosen as $\beta_\gamma = 10^{13}$ and $\beta_m = 10^{2}$.
}
\label{fig:xray_photon_spectrum}
\end{figure*}

Shown in Fig.~\ref{fig:xray_photon_spectrum} is the expected differential photon flux at a distance of 600 km from the surface in the night side of the Earth, for two different choices of $n$ namely, $n=1$ (blue solid line) and $n=3$ (red dotted-dashed line), with $\Lambda = 1\;\mu$eV. The values of the chameleon coupling to photon and matter are same as in Fig.~\ref{fig:solar_chameleon spectrum} - $\beta_\gamma = 10^{13}$ and $\beta_m = 10^{2}$. To calculate this flux, we have used the oscillation probability in Eqn.~\eqref{eqn:osc_prob_Earth} setting $L = 600$ km and using the chameleon mass and photon plasma mass in the Earth's atmosphere at $600$ km to calculate $q$. These have been calculated by considering the mean atmospheric density at $600$ km, which is roughly $1.56 \times 10^{-13}$ kg m$^{-3}$ and the atmospheric electron density at that altitude at night time, which is roughly $7 \times 10^{3}$ cm$^{-3}$~\citep{2022EP&S...74..143L}.

Two things to note from Fig.~\ref{fig:xray_photon_spectrum}. Firstly, the photons produced due to the conversion of chameleons in the geomagnetic field have energies in the range of a few keV meaning that they are X-ray photons. Secondly, we see that for $n=3$ case the X-ray photons have a smooth spectrum whereas for the $n=1$ case the X-ray photon spectrum is suddenly cutoff at around $0.8$ keV. This is because for $n=1$ and the chosen value of $\beta_m$, $m_c (\rho_{\rm core}) \sim 0.8$ keV where $\rho_{\rm core} \simeq 13$ g cm$^{-3}$~\citep{Dziewonski:1981xy}. Hence as discussed previously, chameleons with energies much greater than $0.8$ keV can pass through the Earth whereas those with energies lower than $0.8$ keV are absorbed and hence a sudden drop in the X-ray photon spectrum.

An important point to mention here is that apart from chameleons, axions can also be produced inside the Sun\cite{Raffelt:1985nk} and in the same way as discussed above, they can also result in a X-ray signal in the night side of the Earth due to axion-photon oscillation~\cite{Davoudiasl:2005nh}. However there is a crucial difference between the signal generated by axions oscillating into photons and chameleons oscillating into photons in the geomagnetic field. The photons produced from axions are polarized parallel to the geomagnetic field whereas the photons produced from chameleons are polarized perpendicular to the geomagnetic field. This difference in polarization state of the X-ray photons can be used to distinguish between the X-ray signal from a chameleon and that from an axion.

\begin{figure*}[ht!]
\centering
\includegraphics[width=8cm]{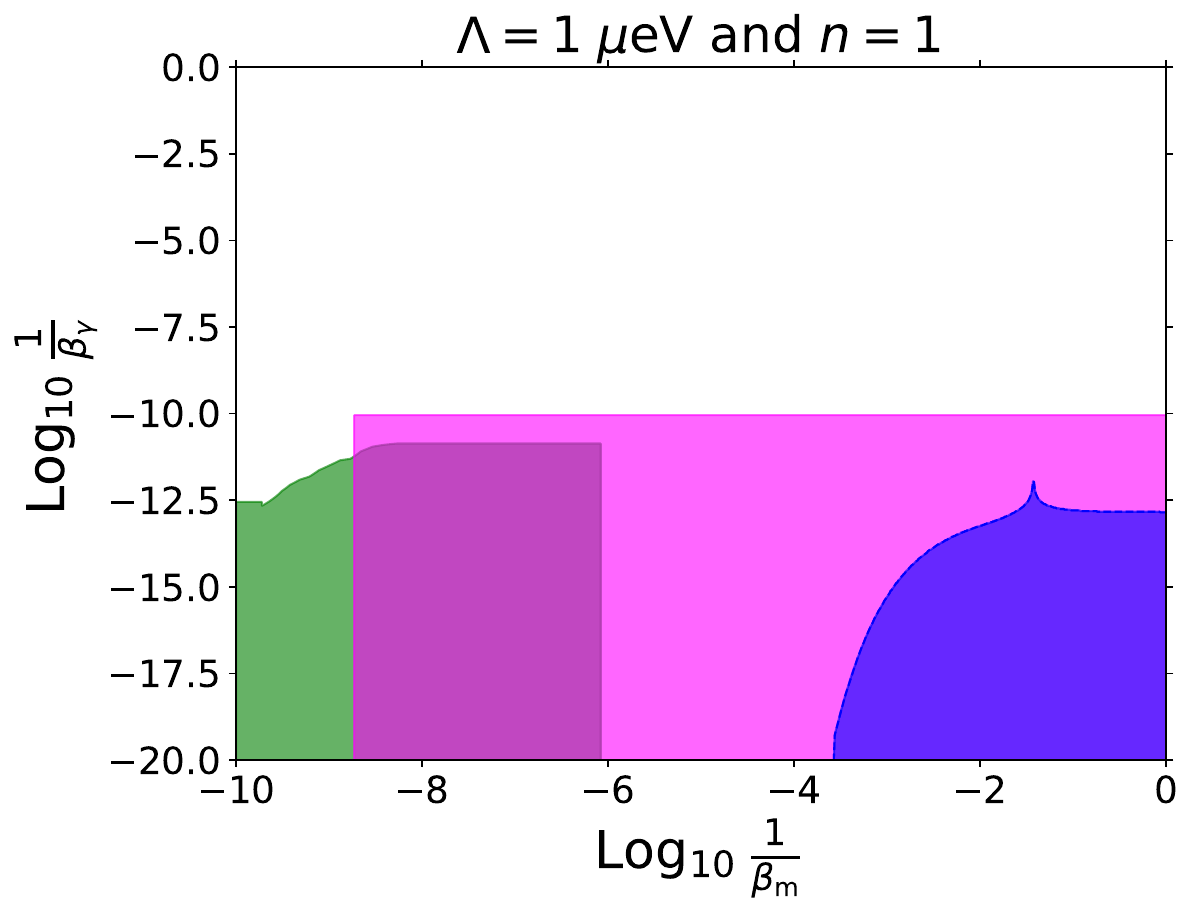}
\includegraphics[width=8cm]{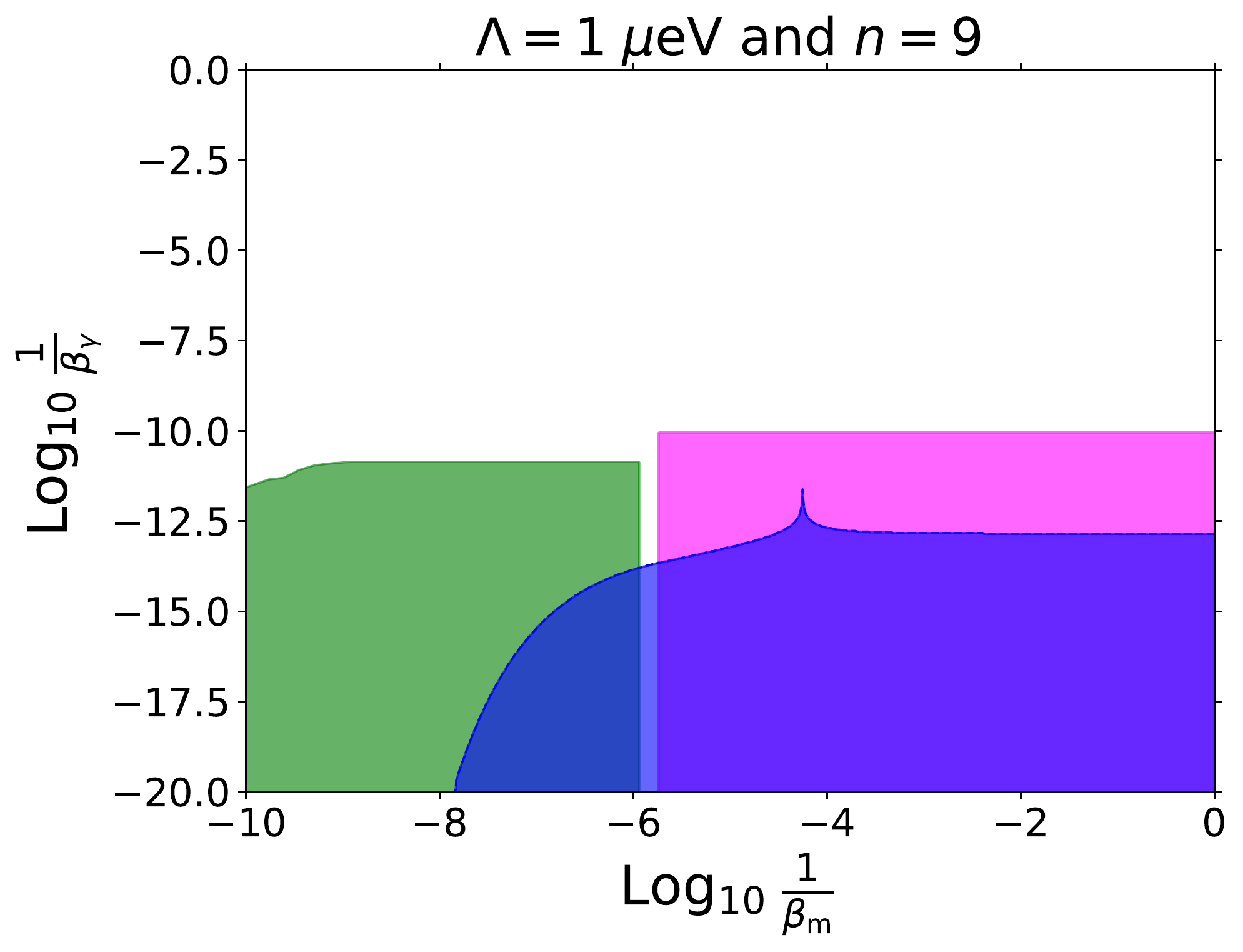}
\includegraphics[width=8cm]{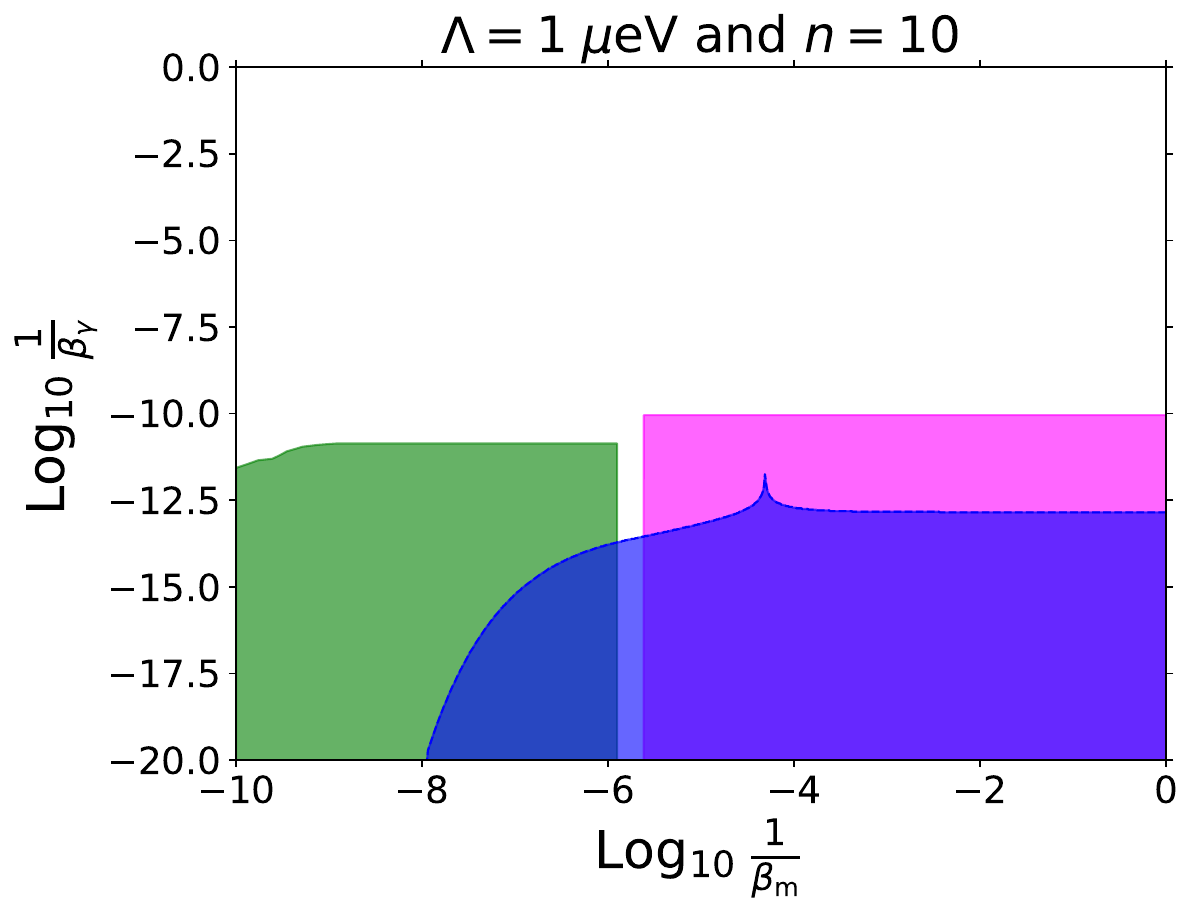}
\includegraphics[width=8cm]{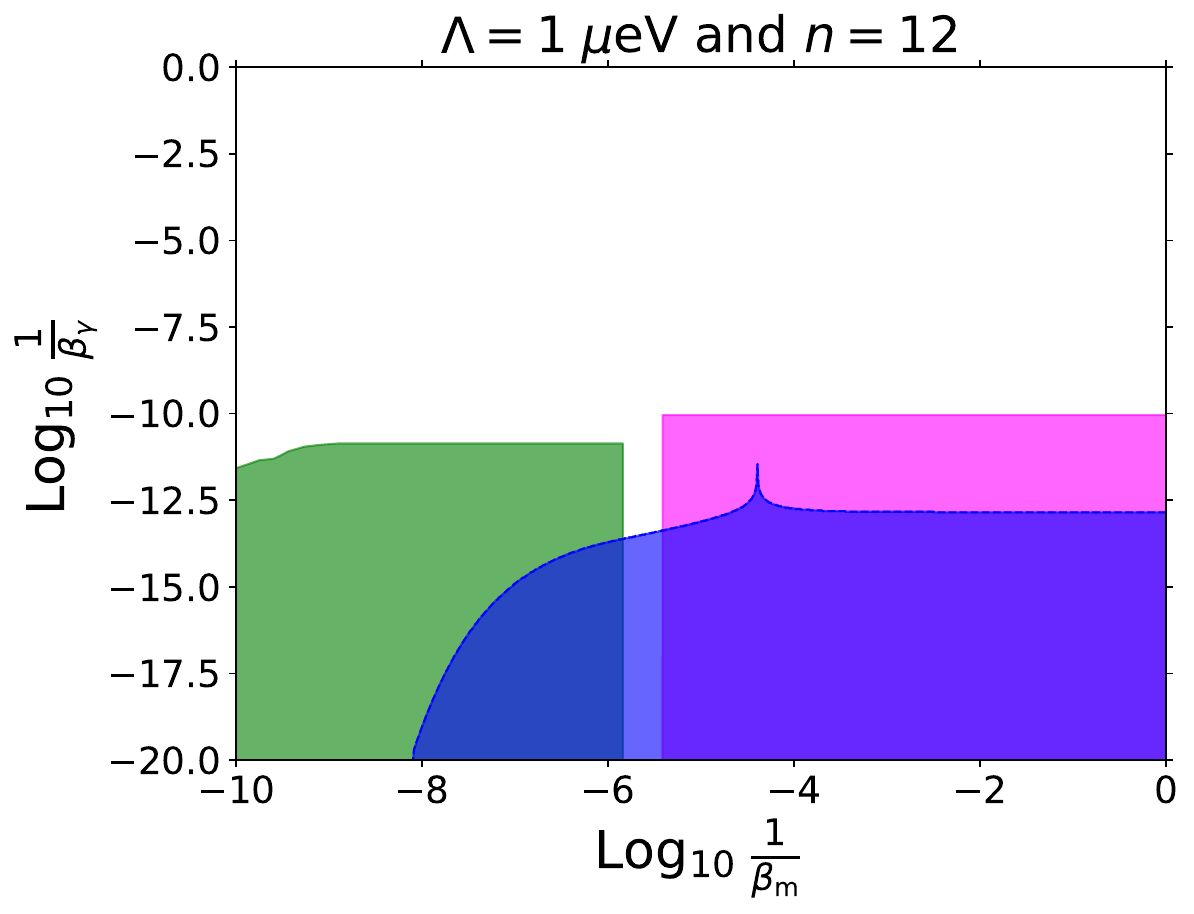}
\caption{Region of the $\beta_m$-$\beta_\gamma$ parameter space of different chameleon models that can be explored by 1 year observation of the night side of Earth by XRISM (blue shaded region). The parameters of the chameleon models are mentioned at the top of each figure. The sharp peak in our constraints is due resonant chameleon production in the solar tachocline leading to a higher chameleon flux and consequently a stronger X-ray signal. Existing constraints from GammeV-CHASE and starlight polarization in this parameter space are shown by the green and pink shaded regions respectively. 
}
\label{fig:chameleon_constraints}
\end{figure*}

The X-ray photons thus produced can be detected by an X-ray telescope orbiting the Earth and passing through the night side. As an example in this work we study the potential of XRISM, an upcoming space based X-ray telescope, to detect this X-ray signal.

XRISM~\citep{XRISM:2020rvx} is a next generation X-ray telescope launched in September 2023 as a successor to the erstwhile Hitomi X-ray telescope. XRISM is placed at a low Earth orbit at an altitude of 550 km. It is currently undergoing calibration and is expected to make observations soon. The XRISM is equipped with two instruments for studying soft X-ray - Resolve X-ray spectrometer and Xtend X-ray imager. The Resolve instrument will have similar capability as that of the Soft X-ray telescope for Spectrometer on board Hitomi whereas Xtend will be an improved version of the Soft X-ray telescope for Imager of the Hitomi mission. In this work we present the possibility of detecting the chameleon-induced X-ray photons using the Resolve instrument of XRISM as an example.

The Resolve instrument is expected to have a background rate of $\leq 2 \times 10^{-3}$ photons s$^{-1}$ cm$^{-2}$ keV$^{-1}$ over an energy range $0.3 - 12.0$ keV and an effective area of approximately $200$ cm$^{2}$~\citep{XRISM:2022exn}. Using the upper bound of the background rate and following~\citep{Davoudiasl:2005nh} we calculate the sensitivity of Resolve over the operating energy range, assuming a uniform background rate over the entire energy range, for 1 year of observation of the night side of the Earth. This turns out to be approximately $7.07 \times 10^{-8}$ photons s$^{-1}$ cm$^{-2}$ keV$^{-1}$. Hence for a given chameleon model we consider the X-ray signal produced in the geomagnetic field to be observable if $\Phi_{\gamma} (\omega)$ is greater than the above mentioned sensitivity for the given values of $\beta_m$ and $\beta_\gamma$. Using this criterion we identify the regions of the $\beta_m$ vs $\beta_\gamma$ parameter space of four different chameleon models - $\Lambda = 1\;\mu$eV and $n=1$, $\Lambda = 1\;\mu$eV and $n=9$, $\Lambda = 1\;\mu$eV and $n=10$ \& $\Lambda = 1\;\mu$eV and $n=12$ - which can produce a detectable X-ray signal.

The blue shaded regions shown in Fig.~\ref{fig:chameleon_constraints} are the regions of the $\beta_m$ vs $\beta_\gamma$ parameter space that can be explored by 1 year of observation of the night side of the Earth by XRISM Resolve for four different chameleon models - $\Lambda = 1\;\mu$eV and $n=1$ (top left), $\Lambda = 1\;\mu$eV and $n=9$ (top right), $\Lambda = 1\;\mu$eV and $n=10$ (bottom left) \& $\Lambda = 1\;\mu$eV and $n=12$ (bottom right). Also shown in the figures are the existing constraints in this parameter space. The green shaded regions are excluded by the GammeV-CHASE experiment~\citep{Steffen:2010ze}. The pink shaded regions are constrained by the non-observation of polarization of light coming from distant stars due to chameleon-photon mixing in the Milky Way magnetic field~\citep{Burrage:2008ii}.

From Fig.~\ref{fig:chameleon_constraints} we see that for the chameleon model $\Lambda = 1\;\mu$eV and $n=1$ (top left) the region of the parameter space where detectable X-ray signals can be obtained is already ruled out by the constraint from starlight polarization. On the other hand for $n=9,10 \; \text{and} \; 12$ models there exists a thin region of the parameter space in between the existing constraints from GammeV-CHASE experiment (green shaded region) and starlight polarization observation (pink shaded region) which can result in an observable X-ray signal. Non-observation of any X-ray signal will consequently rule out this thin blue shaded regions of the three chameleon models, thereby filling the gap between the two existing constraints. For all the other chameleon models between $n=1$ and $n=9$ we have checked that the situation is similar to the $n=1$ case i.e., the region of the parameter space which can produce observable signal is well within the presently excluded region and hence these models are not shown here. On the other hand for chameleon models with $n = 11,13,14 \; \text{and} \; 15$ we found that the situation is similar to the $n = 9, 10 \; \text{and} \; 12$ cases viz., there exists a thin region of the parameter space in between the two existing constraints which can produce observable X-ray signals and can thus be probed using the above setup. Furthermore we observed that with increasing $n$, the width of this thin region increases. Thus for $n \geq 9$ the constraint on the chameleon parameter space from our setup can bridge the gap between the two existing constraints.

\begin{figure*}[t]
\centering
\includegraphics[width=10cm]{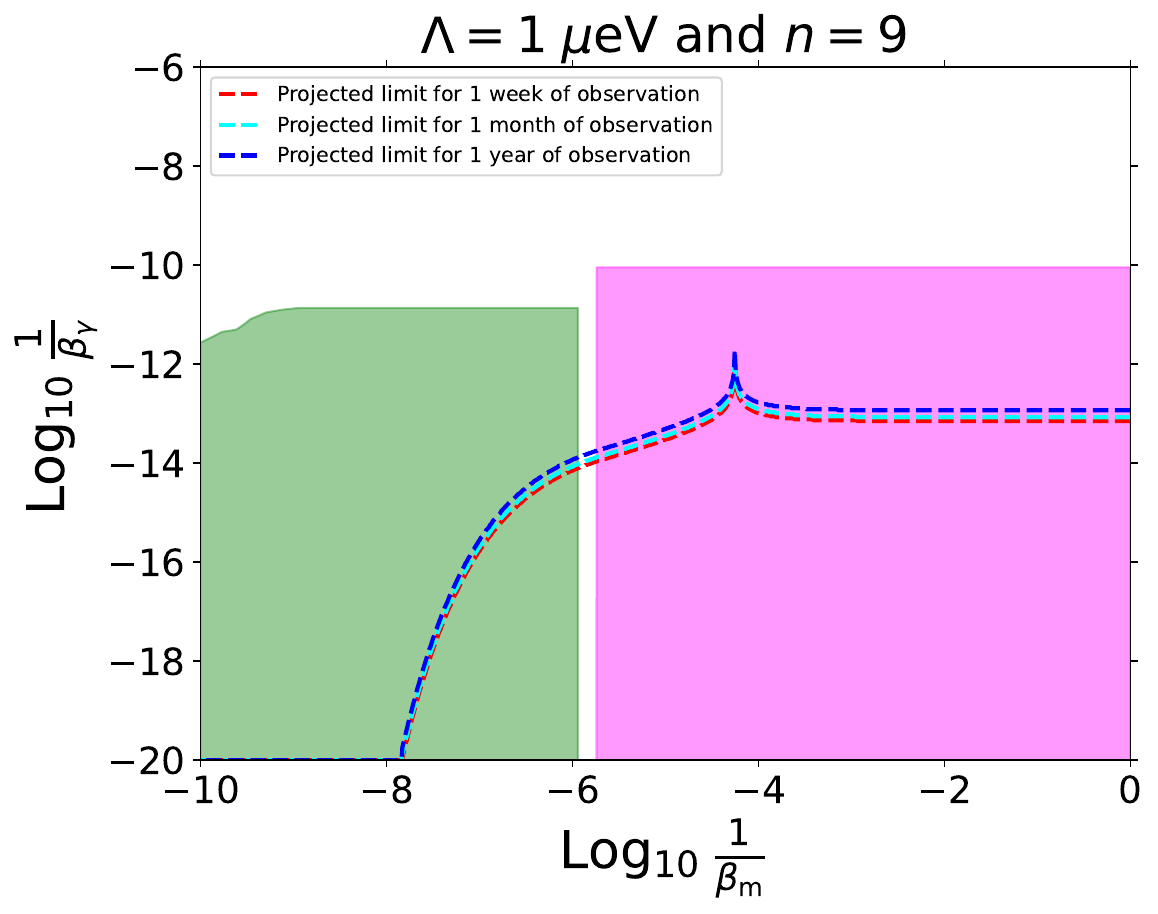}
\caption{XRISM projected limits on the $\beta_m$-$\beta_\gamma$ parameter space for the $\Lambda=1\;\mu$eV and $n=9$ chameleon model for 1 month (cyan dashed line) and 1 week (red dashed line) of observation of the night side of Earth. Also shown for comparison is the projection for 1 year (blue dashed line) of observation of the night side of Earth (same as Fig.~\ref{fig:chameleon_constraints}). Existing constraints from GammeV-CHASE and starlight polarization in this parameter space are shown by the green and pink shaded regions respectively.
}
\label{fig:varying_obs_time}
\end{figure*}

Finally, we show in Fig.~\ref{fig:varying_obs_time} how the region of the $\beta_m$ vs. $\beta_\gamma$ parameter space for the $\Lambda=1\;\mu$eV and $n=9$ chameleon model that XRISM can probe, changes for a shorter (and thereby more feasible) observation time. We find that for one week (month) of observation of the night side of the Earth, XRISM can probe the entire region enclosed by the red (cyan) dashed line in Fig.~\ref{fig:varying_obs_time}. For reference, the region accessible to XRISM for one year of observation is the region enclosed by the blue dashed line (which is the same as shown in Fig.~\ref{fig:chameleon_constraints}). Thus, we find that for shorter observation times, the XRISM can still probe regions that are complementary to the existing constraints on this parameter space. This also turns out to be the case for all chameleon models with $n \geq 9$.

\section{Summary \& Conclusion}
\label{sec:conclu}
To summarize, in this work we have studied the possibility of detecting solar chameleons using the Earth's magnetic field. We have studied chameleon models with $\Lambda = 1\;\mu$eV which are also expected to be detectable by electron-recoil dark matter detection experiments. Consequently we have first shown that chameleons can be produced abundantly in solar tachocline region due to the conversion of thermal photons into chameleons under the influence of the strong magnetic field in the tachocline region. The produced chameleons which have energies greater than their mass in the tachocline can easily escape the Sun thereby resulting in a flux of solar chameleons. We have calculated the differential solar chameleon flux coming out of the Sun and we have shown that this differential flux is peaked in the keV energy range. These escaping chameleons travel through space and reach Earth along with the photons emitted by the Sun.

Once they reach the Earth, all the photons are blocked by the Earth.
However this is not the case with the chameleons. Owing to their weak interaction with matter some of the chameleons can pass through the Earth. This is because, due to their density dependent mass, chameleons with energies less than their mass inside the Earth will be blocked by the Earth. On the other hand chameleons with energies much higher than their mass inside the Earth can pass right through the Earth and emerge out in the night side of the Earth. Once outside these chameleons travel through the Earth's magnetic field and oscillate into photons and back. Since the chameleons typically have energies of the order of few keV, the produced photons give rise to an X-ray signal in the atmosphere.

We show that for certain chameleon models, these X-ray signals are within the reach of future space based X-ray observatories such as XRISM orbiting the Earth. Furthermore this chameleon induced X-ray signal will be polarized parallel to the direction of the Earth's magnetic field. This fixed polarization state is an important feature that can help us uniquely identify whether a potential signal is produced by a chameleon or some other exotic particle such as an axion. Ultimately we show that for the chameleon models considered here, this setup with an orbiting X-ray telescope such as XRISM can explore such regions of the chameleon coupling parameter space which are unconstrained by other experiments.

Although we have derived our projections assuming one year of observation of the night side of the Earth with the XRISM, we find that even for a shorter observation time such as one month or one week, our proposed setup can still place constraints on such regions of the chameleon parameter space which are complementary to the existing constraints. Additionally we find that on increasing the observation time of the night side of the Earth, our limits on $\beta_\gamma$ (blue shaded region) also improve. Nevertheless we have found that in order to surpass the existing limits (pink and green shaded regions) with this setup and probe even smaller values of $\beta_\gamma$, one would need an abnormally large observation time. An alternate way to probe such smaller values of $\beta_\gamma$ (and probably the more practical way) would be to use a dedicated X-ray detector with a much lower background rate which can achieve a higher sensitivity with a shorter observation time. Such an experiment would in turn enable one to place even stronger bounds in $\beta_m$ vs $\beta_\gamma$ parameter space.

We would like to conclude by mentioning that, apart from chameleons, another popular beyond standard
model particle, namely the axion, which if it exists, can
also produce similar X-ray signals that can be detected by XRISM. Furthermore, the axion-induced X-ray photons will be polarised parallel to the external magnetic field, as opposed to the chameleon-produced X-ray photons, which are polarised perpendicular to the external field. Hence, this natural difference in the polarisation of the X-ray photons can be used to distinguish between the two particles. However, the Resolve instrument of the XRISM telescope is not capable of conducting such polarisation studies of X-rays, and hence it will not be able to distinguish between signals produced by an axion and a chameleon. In order to simultaneously search for solar
axions and solar chameleons, we would require an X-ray telescope that can detect and measure the polarisation of X-rays. Telescopes such as the IXPE~\cite{IXPE} and the recently launched XPoSat~\cite{XPoSat}, which are dedicated polarimetry missions to study X-ray polarisation, might be capable of simultaneously searching for axions and chameleons.
However, such a polarisation study is beyond the scope of the present paper, and we leave it for future work.

\section*{Acknowledgements}
TK acknowledges support in the form of Senior Research Fellowship (File No. 09/0080(13437)/2022-EMR-I) from the Council of Scientific \& Industrial Research (CSIR), Government of India.

\bibliography{refs}

\end{document}